\documentclass{article}

 \PassOptionsToPackage{table}{xcolor}

\usepackage[preprint, nonatbib]{neurips_2025}

\usepackage[utf8]{inputenc} 
\usepackage[T1]{fontenc}    
\usepackage[colorlinks]{hyperref}       
\usepackage{url}            
\usepackage{booktabs}       
\usepackage{amsfonts}       
\usepackage{nicefrac}       
\usepackage{microtype}      
\usepackage{listings}
\usepackage{todonotes}
\usepackage{tikz}
\usetikzlibrary{arrows.meta, positioning, shapes.geometric}
\usepackage{xspace}
\usepackage{listings}
\usepackage{float}
\usetikzlibrary{calc}
\usepackage{subcaption}
\usepackage{paralist}
\usepackage{comment}
\usepackage{amsmath}

\newenvironment{paraenum}{\begin{inparaenum}[\itshape (i)\upshape]}{\end{inparaenum}}

\lstdefinestyle{Python}{
  language=Python,
  basicstyle=\ttfamily\small,
  keywordstyle=\bfseries\color{blue},
  commentstyle=\itshape\color{gray},
  stringstyle=\color{red},
  showstringspaces=false,
  numbers=none,
  breaklines=true,            
  morekeywords={frm, let, cls, fun, act},
}
\title{Rulebook: bringing co-routines to reinforcement learning environments}

\author{%
  Massimo Fioravanti, Samuele Pasini, Giovanni Agosta\\
  DEIB -- Politecnico di Milano\\
  Via G. Ponzio 34/5, Milano, Italy\\
  \texttt{massimo.fioravanti@polimi.it, samuele.pasini@mail.polimi.it},\\ \texttt{giovanni.agosta@polimi.it} \\
}

\newcommand{\Rulebook}{\emph{Rulebook}\xspace}
\newcommand{\OpenSpiel}{\emph{OpenSpiel}\xspace}
\newcommand{\CPP}{\texttt{C++}\xspace}
\newcommand{\RLC}{\textsc{rlc}\xspace}
\newcommand{\ActionFunction}{\texttt{action}\xspace}
\newcommand{\ActionFunctions}{\texttt{action}s\xspace}
\newcommand{\gym}{\texttt{gym}\xspace}
\newcommand{\llvm}{\textsc{llvm}\xspace}
\newcommand{\MLIR}{\textsc{MLIR}\xspace}
\newcommand{\TicTacToe}{\emph{Tic-Tac-Toe}\xspace}

\begin{document}

\maketitle

\begin{abstract}
Reinforcement learning (RL) algorithms, due to their reliance on external systems to learn from, require digital environments (e.g., simulators) with very simple interfaces, which in turn constrain significantly the implementation of such environments.
In particular, these environments are implemented either as separate processes or as state machines, leading to synchronization and communication overheads in the first case, and to unstructured programming in the second. 

We propose a new domain-specific, co-routine-based, compiled language, called \Rulebook, designed to automatically generate the state machine required to interact with machine learning (ML) algorithms and similar applications, with no performance overhead. \Rulebook allows users to express programs without needing to be aware of the specific interface required by the ML components. By decoupling the execution model of the program from the syntactical encoding of the program, and thus without the need for manual state management, \Rulebook allows to create larger and more sophisticated environments at a lower development cost.
\end{abstract}

\section{Introduction}
The explosion of machine learning (ML) as a key technique in many domains has driven the creation of many tools and frameworks to build, deploy, and integrate with other systems ML algorithms.
Such tools have mostly been built as libraries rather than domain specific languages (DSLs) -- a perfectly reasonable choice, given that languages require compilers, and compilers have notoriously long development times and are costly to maintain~\cite{leupers2001retargetable}. 
Recently, however, \MLIR~\cite{lattner2020mlir}, released under Apache License v2.0 with LLVM Exceptions, has emerged as a game-changer in this scenario, offering a modular way to build compilers at a fraction of the usual cost, for domain-specific languages.
The advantage provided by \MLIR is such that, after showcasing a full compilation toolchain for the ONNX format~\cite{jin2020compiling}, most neural network frameworks, such as Tensorflow and Pytorch, have started turning to \MLIR as a tool to simplify the deployment~\cite{kwon2022aiwarek,lattner2021mlir}.
These tools focus on the deployment of ML models on hardware platforms, but little effort has been devoted  to a different, but also critical, issue -- that of efficiently and effectively interfacing ML models with other software systems.

Reinforcement Learning (RL) is widely adopted for training agents in scenarios where, rather than a large labeled dataset, the agent can learn from interactions with a physical or simulated system, which is the case in a wide range of applications, from cyber-physical systems to autonomous driving and traffic management to games such as chess or Go~\cite{puiutta2020explainable,gu2024reviewsafereinforcementlearning}.
It is worth noting that even in cyber-physical systems, it is often more practical to perform the learning phase on a simulation model of the physical systems, which enables the exploration of dangerous corner cases (e.g., car accidents) and vastly reduces costs and learning times.
In this scenario, given a RL library and a simulation library, most of the work for the developer consists of interfacing the two.
RL libraries need to interact with a very simple interface, where the RL and the simulation take turns.
In their respective turns, the RL selects an action based on the simulation state and its own knowledge, while the simulation evaluates the evolution of the system based on the input action received from the RL library.
In principle, this interaction can be modeled by each actor (RL and simulation) running in its own process or thread, and waiting for the other to take its turn, with communication performed via messages or shared memory.
However, this method imposes significant communication and synchronization overheads, since it parallelizes activities that are actually necessarily sequential.
The alternative is to weave the control flow of a single process so that it switches between the two libraries.
This, however, generates uselessly complex control flow.
A possible solution is provided by cooperative multi-tasking, implemented via co-routines. 
While this solution provides a simple way to alternate between the two actors with minimal overheads, there are more subtle limitations, tied to the difficulty of inspecting, serializing, or otherwise interacting with the co-routine state, which currently prevent its widespread used in the field of RL.

In this paper, we propose \Rulebook, a DSL that enables inspection, serialization and de-serialization of co-routines, simplifying the collection of observations for the RL agents.
We implement the \Rulebook compiler, \RLC, through the \MLIR compiler construction framework, which allowed us to implement a fully working, high-performance solution in less than 2 person-years (a major advantage compared to the hundreds of person-years needed to develop a compiler from scratch), and test it against the \OpenSpiel implementation of a set of well-known board games to prove its effectiveness and efficiency at enabling the learning of playing strategies for such games.
We measure the relative speed of \Rulebook ranging from 0.81x to 3.49x when compared to \OpenSpiel implementations. Furthermore, we provide the implementation of a large commercial off-the-shelf board game as a example of \Rulebook production-readiness.  

The rest of this paper is organized as follows.
In section~\ref{sec:bg} we discuss common practices to implement RL, and highlight the limitations which justify the introduction of the \Rulebook language.
In section~\ref{sec:sol}, we describe the design goals of \Rulebook and the language features that enable to achieve them, while in section~\ref{sec:eval} we provide an experimental assessment of \Rulebook and its compiler \RLC, comparing on RL tasks against \CPP implementations of board games from the \OpenSpiel suite.
Finally, in section~\ref{sec:conc}, we draw some conclusions and highlight future research directions.

\section{Background}
\label{sec:bg}
RL studies how intelligent agents ought to take actions when they exist in a dynamic \emph{environment}, which, in our case, are digitally simulated.
The field of RL has coalesced around system designs where the ML components are, from a software engineering perspective, fully separated from the environment they interact with. 
These systems often use libraries such as \gym \cite{gym} or \texttt{gymnasium} \cite{kwiatkowski2024gymnasium}, which achieve the separation of the RL algorithm from the environment by wrapping the latter in a simple interface object, as demonstrated in listing \ref{lst:env} for the case of a \TicTacToe game.
Such interfaces usually consist of three methods: 
\begin{paraenum}
\item An \lstinline[style=Python]{__init__} method that sets up the environment so that information about the dimensions of the observations, the state space, and so on can be accessed by users of the class. 
\item A \lstinline[style=Python]{reset} method that restores the initial state of the simulation and returns the initial observation. 
\item A \lstinline[style=Python]{step} method that accepts a description of the action to be taken (the ID of the \TicTacToe cell in this case), executes the action, and returns the new observations. 
\end{paraenum} 

\begin{figure}[htbp]
\captionof{lstlisting}{\gym \TicTacToe wrapper example}
\begin{minipage}[t]{0.45\textwidth}
\begin{lstlisting}[style=Python, label=lst:env]
class TicTacToe(gym.Env):
  def __init__(self):
    self.observation_space = 
      gym.spaces.Box([0, 0, 0], [3, 3, 3])
    self.action_space = gym.spaces.Discrete(9)
    ... # set up game data
\end{lstlisting}    
\end{minipage}\hfill
\begin{minipage}[t]{0.45\textwidth}
\begin{lstlisting}[style=Python, firstnumber=last]

  def reset(self):
    ... # reset game data
    return self._observation()

  def step(self, cell_id):
     ... # mark cell at cell_id 
     return self._observation()
\end{lstlisting}
\end{minipage}
\end{figure}

\paragraph{Asynchronous vs synchronous environments}
The execution of the environment code can run in a separate thread or process from the wrapper (\emph{asynchronous} implementation),  or it can run in the same thread as the wrapper object (\emph{synchronous} implementation). 
The asynchronous implementations allow more freedom in the implementation itself, as long as the communication is handled correctly, but impose undesirable communication and synchronization overheads.
The synchronous implementations must yield control back to the caller of the wrapper, allowing a new step to be triggered according to its logic, and may do so by manually unfolding the state machine implied in the asynchronous version, as shown in listing \ref{lst:s_tic_tac_toe} for \TicTacToe.

\begin{figure}[htbp]

\begin{lstlisting}[style=Python, label=lst:s_tic_tac_toe, caption={Sketch of the synchronous implementation of \TicTacToe}]
class TicTacToe:
  def __init__(self):
    self.board = TicTacToeBoard()
    self.current_player, self.winner = 0, None
    self.next_resumption_point = NormalTurn

  def is_done(self) -> Bool:
    return self.next_resumption_point == Ended

  def can_mark(self, x: int, int: y) -> Bool:
    return self.next_resumption_point == NormalTurn and x >= 0 and 
       x < 3 and y >= 3 and y <= 3 and not self.board.is_set(x, y)

  def mark(self, x: int, y: int):
    assert self.can_mark(x, y)
    self.board.set_marked_by_current_player(x, y)
    if self.board.three_in_a_line():
      self.next_resumption_point = Ended
      self.winner = self.current_player
    else: self.current_player = (self.current_player + 1) % 2
\end{lstlisting}
\end{figure}

Here \lstinline[style=Python]{next_resumption_point} keeps track of the state of the game, allowing to suspend the execution of the game after each action and yield control back to the caller. It it worth noting that many ML projects adopt this approach \cite{koyamada2023pgx,LanctotEtAl2019OpenSpiel,DBLP:journals/corr/abs-1911-08265,muzero-general,ye2021mastering}.
The conversion between the \emph{asynchronous} and \emph{synchronous} implementation is similar to the conversion of a higher level programming language to assembly code executed by a compiler. High level control flow structures are lowered into blocks of code that connect to each other through jumps and conditional jumps, which manipulate the program counter.
In most scenarios manual management of jumps is considered bad practice \cite{consideredharmfull}.

\paragraph{Co-routines and their limitations}
Various programming languages attempt to solve this issue by providing solutions such as Python or \CPP co-routines. Co-routines have been known to be a mechanism to allow multiple programs to interoperate \cite[p.~194–195]{taocp}. E.g., listing \ref{lst:coro_tictactoe} shows a Python \TicTacToe implementation that uses coroutines to suspend the execution when actions must be taken, and resume once they are.
\begin{figure}[b]
\begin{lstlisting}[style=Python, label=lst:coro_tictactoe, caption={Partial co-routine TicTacToe}]
def coro_tic_tac_toe(connection):
  board = TicTacToeBoard()
  while not board.is_full():
    (x, y) = yield board.to_observation()
    board.mark_slot(x, y)
    if board.someone_has_won():
        connection.send_victory(board.get_current_player())
        return
    board.switch_current_player()
  connection.send_draw()
  return

frame = coro_tic_tac_toe()
first_obs = frame.send(None)
second_obs = frame.send((0, 1))
\end{lstlisting}
\end{figure}
%
\label{sec:bg:limits}
In practice this solutions are not often used in RL due to a number of limitations in the existing implementations.
\subparagraph{Checking preconditions} Co-routine implementations usually have no lexical constructs to express preconditions over eventual arguments required by the co-routine to resume. E.g., in listing \ref{lst:coro_tictactoe} there is no mechanism to ensure that x and y passed by the user are within the bounds of the board, or that the slot selected by the user has not already been filled. 
If is possible simulate a precondition mechanism by rewriting all yield statements as while loop that keep executing the yield until a acceptable input has been provided, but this is a repetitive and error prone task.
\subparagraph{Co-routine variables are inaccessible} Usually, co-routine implementations do not allow to access stack variables declared inside the co-routine from outside the co-routine, unless the use manually saves a reference to them somewhere. E.g., if listing \ref{lst:coro_tictactoe} was rewritten to use \CPP co-routines, it would not be possible to access the variable board defined inside \lstinline[style=Python]{coro_tic_tac_toe} from outside that co-routine. This is not strictly required by machine learning algorithms, but it is useful for other development tasks, such as writing a program that visualizes the environment. 
\subparagraph{Serialization and deserialization} ML algorithms usually need to access a serialized encoding of the environment, often as an array of floats. Common co-routine implementations typically do not allow to automatically serialize and deserialize co-routines. 

\section{Proposed solution: the \Rulebook language}
\label{sec:sol}
We propose \Rulebook, a \emph{MLIR-based} \cite{lattner2020mlir} domain-specific compiled language that allows one to write a program \emph{as if} it were \emph{asynchronous}, but to use it \emph{as if} it were written as a \emph{synchronous} program within the same thread, by automatically generating the underlying state machine. 
As mentioned in section~\ref{sec:bg:limits}, this is already partially achievable in languages such as Python, with syntactical constructs such as \emph{co-routines}, but limitations prevent them for being used in practice. 
\Rulebook generalizes the co-routine mechanism by integrating it more deeply in its typechecking mechanism, thus enabling inspection of the content of a suspended co-routine, serialization and deserialization, as well as ensuring that no heap allocation is triggered in the lifetime of a \Rulebook co-routine, unless requested by the user. 
E.g., a partial implementation of \TicTacToe in \Rulebook is shown in listing \ref{lst:tictactoe}, while listing \ref{lst:tictactoe_use}, written in the same file, shows how to instead use it in a imperative way.
\begin{figure}[htbp]

\noindent\begin{minipage}[t]{.45\textwidth}
\begin{lstlisting}[style=Python, caption={\Rulebook \TicTacToe declaration},label={lst:tictactoe}]
act play() -> TicTacToe:
  frm board : Board
  while !full(board):
    act mark(Int x, Int y) {
      x<3, x>=0, y<3, y>=0,
      board.get(x, y) == 0
    }
    board.set(x, y)
    if board.three_in_a_line():
      return
    board.change_player()
\end{lstlisting}
\end{minipage}\hfill
\begin{minipage}[t]{.45\textwidth}
\begin{lstlisting}[style=Python, caption={\Rulebook \TicTacToe usage},label={lst:tictactoe_use}]
fun main() -> Int:
  let game = play()
  game.mark(0, 0)
  game.mark(1, 0)
  game.mark(1, 1)
  game.mark(2, 0)
  game.mark(2, 2)
  return game.board.three_in_a_line()
\end{lstlisting}
\end{minipage}
\end{figure}

Thus decoupling the program syntactical \emph{synchronicity} from its execution \emph{synchronicity}.
Furthermore, since the typechecking mechanism knows more about \Rulebook \emph{co-routines}, from a single file, it can generate many useful tools, such as fuzzers, C-compatible headers, Python wrappers, human-readable and binary printer and parses, both for the state of a environment, and for a sequence of actions executed in that environment. 
In the rest of this document we will show how this is achieved and what benefits are obtained from doing so. 

\begin{figure}[htbp]
\centering
\begin{tikzpicture}[
  node distance=0.4cm and 1.4cm,
  every node/.style={draw, rectangle, rounded corners, minimum width=3cm, minimum height=0.8cm, align=center, font=\small},
  process/.style={fill=blue!20},
  file/.style={fill=green!20},
  module/.style={fill=orange!20},
  ml/.style={fill=purple!20},
  arrow/.style={-{Stealth}, thick},
]

\node[process, fill=cyan!30] (input) {Input File\\\textbf{tictactoe.rl}};

\node[process, fill=blue!30, below=of input] (compiler) {Compiler\\\RLC};

\draw[arrow] (input) -- (compiler);

\node[file, below left=of compiler] (exe) {Executable\\\textbf{tictactoe.exe}};
\node[file, left=of compiler] (header) {Printing-Parsing tools\\\textbf{helpers.lib}};
\node[file, below right=of compiler] (py) {Python Wrapper\\\textbf{wrapper.py}};
\node[file, right=of compiler] (fuzzer) {Fuzzer Executable\\\textbf{tictactoe-fuzzer.exe}};
\node[file, below=of compiler] (lib) {Library\\\textbf{tictactoe.lib}};

\draw[arrow] (compiler) -- (exe);
\draw[arrow] (compiler) -- (header);
\draw[arrow] (compiler) -- (py);
\draw[arrow] (compiler) -- (fuzzer);
\draw[arrow] (compiler) -- (lib);

\node[module, below=of py] (gym) {\gym Wrapper\\\textbf{gym-wrapper}};
\node[ml, left=of gym] (ml) {Machine Learning\\\textbf{Training}};
\node[file, left=of ml] (network) {Trained Model\\\textbf{network.pt}};

\draw[arrow] (py) -- (gym);
\draw[arrow] (gym) -- (ml);
\draw[arrow] (ml) -- (network);

\node[draw, dashed, inner sep=0.5cm, label=below right:{}] {};
\end{tikzpicture}
\caption{\RLC toolchain and generated tools}
\end{figure}
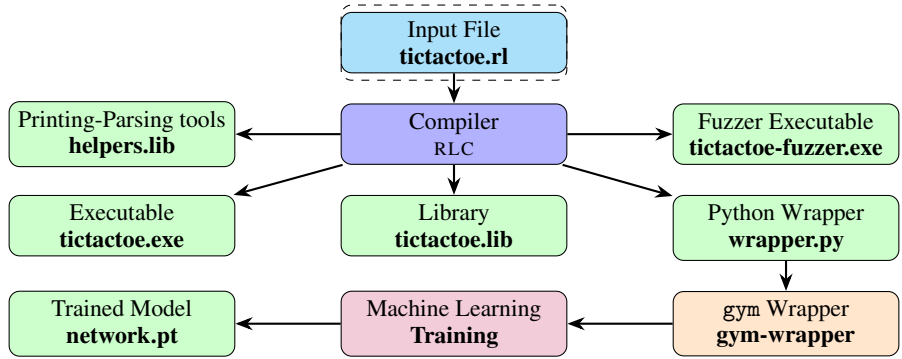

\subsection{\Rulebook Design Goals and Language features}
To be useful for ML purposes, \Rulebook supports the following requirements through its language features: 
\begin{paraenum}
    \item Main-loop Independence;
    \item Inspectability;
    \item Serializability;
    \item Checkability;
    \item C Compatibility.
\end{paraenum}

\paragraph{Main-loop Independence} 
\emph{Environments} should be written as if they had control over the main loop of the program, yet they should be executed as if they did not. 
Main-loop independence is achieved through the implementation of stackful co-routines.

\paragraph{Inspectability} 
While some \emph{environments} exist as ML training tools only, in some use cases the digital \emph{environment} can be used for other purposes (e.g., providing the implementation logic for a computer-based game). 
Thus, the content of a \Rulebook environment should behave as a regular data structure, which can be inspected and modified through code.
Typical co-routines implementations do not allow to inspect the co-routine local variables. 
E.g., \CPP co-routines, when first created, always return a \texttt{std::coroutine\_handle} object, which is not specific to the particular co-routine and does not allow to inspect the co-routine object, only to cancel it or resume it.

\Rulebook instead fully specifies how the co-routine state is laid out in memory, therefore allowing to produce a class declaration that describes the co-routine state object.
F.i., the top level declaration \lstinline[style=Python]{act play() -> TicTacToe:} in listing~\ref{lst:tictactoe} is an \ActionFunction declaration. 
The \lstinline[style=Python]{TicTacToe} return type is declared by the presence of its name in the trailing return type of an \ActionFunction, such as the one in the example. While a file is being type-checked, each \ActionFunction is inspected and, in our example, the class and function shown in listing \ref{lst:tictactoecls} are generated.

\begin{figure}[htbp]
\centering
\captionof{lstlisting}{Class generated by the inspection of the tic tac toe \ActionFunction declaration}
\begin{minipage}[t]{0.45\textwidth}
\begin{lstlisting}[style=Python, label=lst:tictactoecls]
cls TicTacToe:
  Int resume_idx
  # extra fields to be discovered 
  ...
  fun is_done() -> Bool:
   return self.resume_idx == -1
\end{lstlisting}
\end{minipage}\hfill
\begin{minipage}[t]{0.45\textwidth}
\begin{lstlisting}[style=Python,firstnumber=last]
# notice: obtained from, 
# but semantically distict 
# from act play() -> TicTacToe
fun play() -> TicTacToe: 
  let coro_state : TicTacToe
  coro_state.resume_idx = 0
  return coro_state
\end{lstlisting}
\end{minipage}
\end{figure}

The extra fields inside \lstinline[style=Python]{cls TicTacToe} are composed by all variables declared inside the body of the \ActionFunction play marked with the \lstinline[style=Python]{frm} qualifier, such as \lstinline[style=Python]{frm board : Board}.
%
The \lstinline[style=Python]{frm} keyword means \emph{frame}, it is only accepted within \ActionFunction declarations, and means that the lifetime is bound to the frame of the co-routine, and not the lifetime of the current invocation of the co-routine. Since such variables are now accessible trough the class, this is enough to achieve inspectability.

\paragraph{Serializability} 
\emph{Environments} must be serializable and deserializable, at least in a binary form usable by ML components, and possibly as human-readable text. Serializability is a stronger property w.r.t. inspectability, because data structures containing references may be inspectable, but not serializable.
Serializability is achieved by adopting a zero-overhead \emph{traits} system, similar to \CPP templates or \emph{Rust} traits. This allows user defined types to override their serialization mechanism, without introducing virtual calls, thus enabling arbitrary serialization schemes, should the default one not be enough. E.g., the serialization mechanism that converts a boolean object into a one-hot vector of floats, implemented in the standard library of the language, is shown in listing \ref{lst:serialization}.
\begin{lstlisting}[float,style=Python, label=lst:serialization, caption=Bool serialization function from \RLC standard library]
fun write_in_observation_tensor(Bool obj, Int observer_id, 
                        Vector<Float> output, Int index):
    output[index] = float(obj)
    index = index + 1
\end{lstlisting}

\paragraph{Checkability} 
\emph{Environments} should provide a way to check if each action an agent wishes to perform is valid or not. 
Typical \emph{co-routine} implementations can only be resumed, and there is no way to evaluate before execution whether an action is valid. 
Some tools (e.g., fuzzers), may also require \textbf{enumerability}, i.e., the possibility to enumerate all valid actions that can be executed in a given state. 

To allow users of a \Rulebook \ActionFunction to check if the co-routine implementing such action can be resumed with the parameters provided we modify the \lstinline[style=Python]{yield} statements syntax used by other languages to include preconditions.
Such statement is displayed in the \TicTacToe example, called \textbf{Action Statement}, and is shown in listing \ref{lst:tictactoe} at line 4.
Although syntactically similar, \textbf{Action statements} only exist within an \ActionFunction and have distinct semantics. They represents suspension points similar to Python \lstinline[style=Python]{yield} statements, as well as declaring inside the parentheses which parameters are expected by a caller. Inside the brackets is instead specified which preconditions must be met to resume the co-routine. When such statements are encountered by the compiler, each of them generates 2 member functions inside the type declared by their parent \ActionFunction, one to resume the co-routine, and one to check the precondition. In the \emph{TicTacToe} example, the final declared class, after optimizations, is shown in listing \ref{lst:class}. 
\begin{figure}[htbp]
\captionof{lstlisting}{Example of translated \Rulebook action statement from listing~\ref{lst:tictactoecls}}
\begin{minipage}[t]{0.45\textwidth}
\begin{lstlisting}[style=Python, label=lst:class]
cls TicTacToe:
  Int resume_idx
  Board board

  fun can_mark(Int x, Int y) -> Int:
    return self.resume_idx==0 
       and x < 3 and x >= 0 and 
       y < 3 and y >= 0 and 
       self.board.get(x, y)==0

\end{lstlisting}
\end{minipage}\hfill
\begin{minipage}[t]{0.45\textwidth}
\begin{lstlisting}[style=Python,firstnumber=last]
  fun mark(Int x, Int y)->Int:
    assert(self.can_mark(x, y)) 
    self.board.set(x, y)
    if self.board.three_in_a_line():
      self.resume_idx = -1
      return self.board.
       change_current_player()
        
  fun is_done() -> Bool:
    return self.resume_idx==-1 
\end{lstlisting}
\end{minipage}
\end{figure}
Once this class has been generated, the original \ActionFunction can be discarded, and the compilation can resume as usual. 
Since we fully know the API surface of the class, as well as which methods are valid to be invoked, other tools, such as fuzzers, can be automatically configured to execute \TicTacToe \ActionFunction for their own purposes.

\paragraph{C Compatibility} To enable interoperability with the wider ML ecosystem, compatibility with C is achieved by employing the same ABI to lay out data structures in memory, making \Rulebook interoperable with C and \CPP. 
Furthermore, \Rulebook functions can be invoked from Python code.

\subsection{Extra functional properties}
Besides functional requirements, \Rulebook has been designed to make programs easier to write and maintain than and as fast as equivalent programs in C/\CPP.
Efficiency is obtained by taking advantage of \llvm and \MLIR and making \Rulebook fully compiled.
It is analyzed in full in Section~\ref{sec:eval}.

\paragraph{Static analysis advantage}
We provide an intuitive proof that co-routine-based programs expose more static information to the compiler. In particular, we show that the control flow graph of the co-routine implementation is similar to the state machine equivalent to the regular language of the sequences of valid actions of the environment\cite[p.~83]{regex}. Instead, no such information can be extracted from a \gym-like implementation.

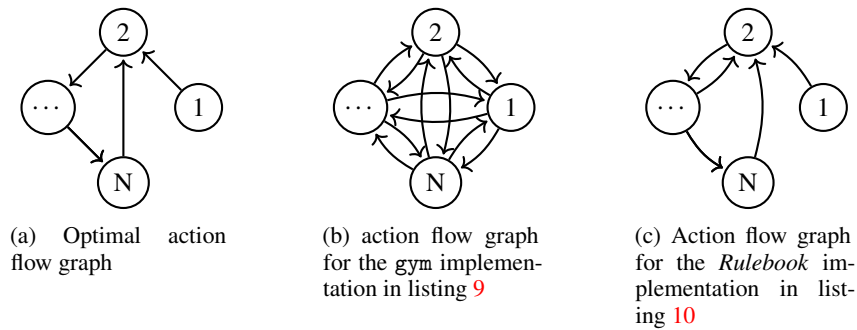
\begin{figure}[ht]
    \centering
    \begin{minipage}{0.8\textwidth}
        \centering
\subcaptionbox{Optimal action flow graph\label{fig:graphs:semantics}}{
\begin{tikzpicture}[shorten >=1pt, auto, node distance=3cm, thick]

\node[circle, draw] (1) at (0:1) {1};
\node[circle, draw] (2) at (90:1) {2};
\node[circle, draw] (3) at (180:1) {\dots};
\node[circle, draw] (4) at (270:1) {N};

\foreach \i/\j in {1/2, 2/3, 3/4, 3/4, 4/2} {
    \path[->] (\i) edge (\j);
}
\end{tikzpicture}
}
\hfill
\subcaptionbox{action flow graph for the \gym implementation in listing~\ref{lst:gym_impl}\label{fig:graphs:gym}}{
\begin{tikzpicture}[shorten >=1pt, auto, node distance=3cm, thick]

\node[circle, draw] (1) at (0:1) {1};
\node[circle, draw] (2) at (90:1) {2};
\node[circle, draw] (3) at (180:1) {\dots};
\node[circle, draw] (4) at (270:1) {N};

\foreach \i/\j in {1/2, 1/3, 1/4, 2/3, 2/4, 3/4} {
    \path[->] (\i) edge[bend left=15] (\j);
    \path[->] (\j) edge[bend left=15] (\i);
}

\end{tikzpicture}
}
\hfill
\subcaptionbox{Action flow graph for the \Rulebook implementation in listing~\ref{lst:rlc_cfg}\label{fig:graphs:coroutine}}{
\begin{tikzpicture}[shorten >=1pt, auto, node distance=3cm, thick]

\node[circle, draw] (1) at (0:1) {1};
\node[circle, draw] (2) at (90:1) {2};
\node[circle, draw] (3) at (180:1) {\dots};
\node[circle, draw] (4) at (270:1) {N};

\foreach \i/\j in {1/2, 2/3, 3/4, 3/4, 4/2, 3/2} {
    \path[->] [bend right=15] (\i) edge (\j);
}

\end{tikzpicture}
}
\end{minipage}
\caption{Action graphs}
\label{fig:graphs}
\end{figure}

Consider an \textbf{Action Flow Graph} (AFG), defined as a directed graph that has a node for each possible action, and an arc between actions $a$ and $b$ when $b$ can be executed after $a$ in a state of the simulation. 
An example of such graph is shown, for an environment with $N$ actions, in figure~\ref{fig:graphs:semantics}.
AFGs, when the nodes associated with actions that can be taken at the start of the environment are annotated as entry nodes and nodes associated with actions that can terminate the environment are annotated as exit nodes, are valid finite-state machines (FSA), equivalent to a regular language that is a super-set of the valid execution traces of the environment\cite[p.~90]{regex}.
In our example, actions $1$ and $N$ are final, and $1$ is the initial action, thus $1(2...N)^*$ is the equivalent regular expression.
This graph is built with human knowledge of the rules of the environment, and is thus optimal -- i.e., as good an approximation as it is possible to achieve with an FSA.
The ability of a compiler to build a conservative approximation of the AFG can be considered as a measure of its ability to perform static analysis on the environment code, and thus to perform other optimizations.


\begin{figure}[htbp]
\captionof{lstlisting}{\gym environment with multiple possible actions}
\begin{minipage}[t]{.43\textwidth}
\begin{lstlisting}[style=Python, label=lst:gym_impl]
class Env(gym.env):
    def __init__(self):
        # setup state
        
    def reset(self):
        # reset state
        return self.obs()
        
    def action1(self, *args):
        # implementation
    ...
    def actionN(self, *args):
        # implementation
\end{lstlisting}    
\end{minipage}\hfill
\begin{minipage}[t]{.5\textwidth}
\begin{lstlisting}[style=Python, firstnumber=last]    

    def step(picked_action):
        if picked_action.id == 1:
            self.action1(*picked_action.args)
        else if picked_action.id == 2:
            ...
        else if picked_action.id == n:
            self.actionN(*picked_action.args)
        return self.obs()
\end{lstlisting}
\end{minipage}
\end{figure}

The \gym environment shown in listing \ref{lst:gym_impl} implements the environment from our example.
The structure of the step function is mandatory. No matter how each action is implemented, for each distinct action there must be at least a check to see if such action is the one to be executed now, and if it is, the correct procedure must be executed. By just inspecting the code, the compiler cannot infer which actions can be executed one after the other.
This implies, when statically analyzing such \gym-like environments, the only AFG that can be inferred is the fully connected graph shown in figure \ref{fig:graphs:gym}. 

In co-routine based implementations, instead, the \emph{control flow graph} of \lstinline[style=Python]{play()} is always an AFG, which is usually not fully connected, and thus a better approximation of the optimal AFG.
Indeed, listing \ref{lst:rlc_cfg} shows a possible implementation in \Rulebook that matches the example AFG. 
Notice that its AFG, when nodes that do not contain actions are collapsed, is identical to the AFG shown in figure \ref{fig:graphs:coroutine}.
In this scenario, the optimal AFG in figure \ref{fig:graphs} implies that \lstinline[style=Python]{args2.condition2} will never be false, a property that might not be possible to deduce from static analysis. 

\begin{figure}[htbp]
\begin{minipage}[t]{0.5\textwidth}
\begin{lstlisting}[style=Python, label=lst:rlc_cfg, caption=Rulebook program matching figure \ref{fig:graphs}]
act play()  -> Game:
 act action1(Args args)
 while args.condition:
  act action2(Args args2)
  ...
  if args2.condition2:
   act actionN(Args argsN)
   args.condition = argsN.condition
\end{lstlisting}    
\end{minipage}\hfill
\begin{minipage}[t]{0.4\textwidth}
\begin{lstlisting}[style=Python, label=lst:mutual_dependency, caption=Rulebook program with mutual dependencies]
act game_1() -> Game1:
    let inner = game_2()
    inner.some_action()

act game_2() -> Game2:
    act some_action()
    let inner = game_1()
\end{lstlisting}
\end{minipage}
\end{figure}

However, since non co-routine based solutions always describe the universal language and thus give no insight on the AFG, the co-routine based implementations is never worse than the alternative.

\subsection{Limitation: No Mutually Recursive \ActionFunctions}

The \Rulebook type-checking mechanism prevents mutually recursive \ActionFunctions.
This is because \ActionFunctions are fully type-checked, and the class generated by the type-checking may be used in other \ActionFunctions and functions. This implies callee \ActionFunctions must be type-checked before the body of caller functions and of \ActionFunctions. 
F.i., consider listing \ref{lst:mutual_dependency}. When the compiler inspects \ActionFunction \lstinline[style=Python]!game\_1! it must decide if the function \lstinline[style=Python]!some\_action! can be invoked on a object of type \lstinline[style=Python]!Game2!. 
To do so, it must have already have type checked the \ActionFunction \lstinline[style=Python]!game\_2!, but \lstinline[style=Python]!game\_2! depends on \lstinline[style=Python]!game\_1!. 

However, \Rulebook type-checking has an important consequence, as it enables the composition of \ActionFunctions which in turn supports the design of complex environments. This is a key feature of the language, which is not available in commonly used languages such as Python. 

\section{Experimental results} 
\label{sec:eval}

\paragraph{LOC and performance} To measure the speed of code compiled by \RLC we compare programs written in \Rulebook against equivalent \CPP handwritten implementations available in \OpenSpiel~\cite{LanctotEtAl2019OpenSpiel}, released under Apache-2.0 license. We have selected 6 games to be compared, shown in \ref{tab:loc}. These games were selected because they are well known and range from very simple games like \textbf{Catch} to the most complex \OpenSpiel game \textbf{Hanabi}.

In table \ref{tab:loc} we report the size in lines of code of the two implementations. \OpenSpiel implementations include hand written serialization schemes for ML and textual, human-readable display, which are automatically generated by \Rulebook.
The \OpenSpiel \textbf{Hanabi} implementation simply wraps the implementation provided by \cite{hanabi-learning-environment}.

\begin{table}[h!]
    \centering
    \caption{\label{tab:loc}Lines of code count of implementations (\OpenSpiel headers ignored)}
    \small
    \begin{tabular}{lcccccc}
        \toprule
        & \textbf{TicTacToe}& \textbf{Hanabi}  & \textbf{Checkers}  & \textbf{Battleship} & \textbf{catch} & \textbf{Connect Four} \\ \midrule
        Our (\Rulebook) & 187& 185 & 278  & 188 & 108 &  222 \\ 
        \OpenSpiel (.CC files) & 218   & 1555  & 574  & 1269 & 194 & 277 \\
        \textbf{Relative Loc} & \cellcolor{green!50} 0.8$\times$ & \cellcolor{green!50} 0.11$\times$ & \cellcolor{green!50} 0.4$\times$  & \cellcolor{green!50} 0.14$\times$ & \cellcolor{green!50} 0.55$\times$ & \cellcolor{green!50} 0.8$\times$ 
     \\ \bottomrule
    \end{tabular}
\end{table}
To show that the languages speed is comparable, \Rulebook programs have been written as a non-expert user would write them. 
\OpenSpiel programs have not been modified, except for standardizing their interface, and do not include expert user optimizations, such as inline-assembly or vectorization.
While it is difficult to provide fair comparisons between programming languages and between different implementations, we attempt to do so by using google-benchmark \cite{google-benchmark}, released under Apache-2.0 license, to only measure the creation of the initial game state and the application of a randomly generated series of valid actions until the final state of the game is reached. We do so to avoid unfairly measuring the time required to find a valid move, which can be greatly influenced by the particular game measured, and by the way it has been implemented.

Even so, some differences still exists:
\begin{paraenum}
    \item \OpenSpiel games keep track of each executed move by placing them in a list of previous actions as a mechanism to implement \textbf{Undo}. Our implementation does not do so, but we tried to level the playing field by artificially replicating the behavior in the \RLC benchmarks by pushing each executed action into a list.
    \item \OpenSpiel encode each possible move into a single integer, and decode its meaning when actions must be executed. Our implementation instead uses a user-defined data structure to encode actions, which requires no decoding. We attempt to make the comparison fair by creating a table of all possible moves of the game and encode the list of random moves as a list of indexes referring to entries of that table.
    \item \OpenSpiel offers more configurable parameters than our implementations, such as battleship board size, number of ships, and so on. Thus less when compiling their implementations less opportunities to perform constant propagation arise. We have not attempted to replicate each possible configuration. 
\end{paraenum}

\paragraph{Experimental setup} The versions of \RLC and \OpenSpiel employed in the experiments are respectively  \href{https://github.com/rl-language/rlc/}{RLC} at commit 6a237f7 and \href{https://github.com/drblallo/open_spiel/}{open\_spiel} at commit 21c5e38. 
The data were collected on a xubuntu 22.04 machine with 6.5.0-44-generic kernel, 13th Gen Intel(R) Core(TM) i9-13900HX CPU, and 2 RAM banks of SODIMM Synchronous 4800 MHz.  The executable for each benchmark is built by the regular \texttt{cmake} build mechanism and placed in the build directory.

\begin{table}[h!]
    \centering
    \caption{\label{tab:speedup}Time required to playout 1024 random valid traces}
    \small
    \begin{tabular}{lcccccc}
        \toprule
        & \textbf{TicTacToe} & \textbf{Hanabi} & \textbf{Checkers}  & \textbf{Battleship}  & \textbf{Catch} & \textbf{Connect Four} \\ \midrule
        Our (w/ action log) & $0.21 ms$ & $1.13 ms$  & $16.9ms$  & $2.5 ms$ & $0.087ms$ & $1.9ms$ \\ 
        \OpenSpiel (\CPP) & $0.17 ms$ & $2.23 ms$  & $59 ms$  & $473ms$ & $0.072ms$ & $2ms$ \\ 
        \textbf{Speedup} & \cellcolor{red!50} 0.81$\times$ & \cellcolor{green!50} 1.9$\times$ & \cellcolor{green!50} 3.49$\times$  & \cellcolor{green!50} 188$\times$ & \cellcolor{green!50} 1.2$\times$ & \cellcolor{green!50} 1.05$\times$ \\
        Our (w/out action log) & $0.15 ms$ & $1.06 ms$ & $16.7 ms$   & $2.3 ms$ & $0.029ms$ & $1.9ms$  \\ 
        \bottomrule
    \end{tabular}
\end{table}

\paragraph{Analysis}
The \TicTacToe implementation very similar in both languages, and indeed shows that the performances are comparable. When the artificial benchmarking slowdown to account for \OpenSpiel \textbf{Undo} feature and action decoding mechanism are removed, \Rulebook \TicTacToe performs faster than the \OpenSpiel implementation.
The \Rulebook \emph{Battleship} implementation is significantly faster, this is due to the \OpenSpiel implementation prioritizing the ability of performing \textbf{Undo} over the raw action application speed.
Finally, \emph{Checkers} and \emph{Hanabi} speed difference is explained by the \OpenSpiel implementation being more configurable. 
Overall, the result show that binaries generated by \RLC have performances comparable to \CPP code that has not been optimized by an expert.

\paragraph{A case study for larger games} Beside the above-reported benchmarks, with \Rulebook we have also implemented\footnote{ \href{https://github.com/rl-language/4Hammer/}{https://github.com/rl-language/4Hammer}} one of the game modes of \textbf{Warhammer 40,000}, a game with significant rules complexity (weight of 4.23/5 \cite{bbg}). In addition to being the first computer implementation ever produced of such game, it can run in browser, interoperate with \CPP and \textbf{Python}, and the core game rules (that is, ignoring graphical code and game pieces constants) only required $\sim$2500 lines of \Rulebook code, and $\sim$5000 lines of code in total. Beside proving the robustness of RLC, it shows how the co-routine approach helps to manage the combinatorial explosion of rules interactions.

\section{Conclusions}
\label{sec:conc}
We have introduced \Rulebook, a new compiled domain specific language supporting serializable and inspectable co-routines, which enable easy integration of reinforcement learning libraries with simulation environments. 
To show the effectiveness of the approach, we have re-implemented in \Rulebook six boardgames from the \OpenSpiel suite, and proved that \Rulebook code is as fast as equivalent \CPP code, while providing the required inspectability, checkability, serializability, main-loop independence, as well as C compatibility and improving cyclomatic complexity.

Beside merely extending the language and the standard library of the language, \Rulebook is open to both theoretical and practical endeavors.  
In particular, type-checking mechanisms in the \Rulebook language could be explored and expanded to lift the limitations over mutually dependent and recursive \textbf{Action functions}.
\Rulebook also offers a lower development cost over other languages by allowing programmers to write code that more closely resembles the high-level description of sequential decisions making programs, and by automatically generating useful utilities, making it a good candidate for producing RL datasets.

\newpage

\begin{ack}
This work is partially supported by the Italian Ministry of Enterprises and Made in Italy (MIMIT)
under the program “Accordi per l’innovazione nella filiera del settore automotive”, through the grant
"Piattaforma ed ecosistema cooperativo, C- ITS ETSI standard per la mobilità digitale
integrata", numero F/340043/01-04/X59, CUP B49J24001210005, finanziato a valere del Bando
MISE – ACCORDI PER L’INNOVAZIONE NEL SETTORE AUTOMOTIVE D.M. 31/12/2021 e DD
10/10/2022
\end{ack}

\bibliographystyle{unsrt}
\bibliography{bibliography}

\end{document}